\documentclass[twocolumn,floats,prl,aps]{revtex4}

\usepackage{graphicx}

\newcommand\beq{\begin{equation}}
\newcommand\eeq{\end{equation}}
\newcommand\bea{\begin{eqnarray}}
\newcommand\eea{\end{eqnarray}}
\newcommand{\nonum}{\nonumber}

\begin{document}

\title{\bf Length-Scale Dependent Superconductor-Insulator Quantum Phase Transition
in One Dimension} 

\author{\bf Sujit Sarkar}
\address{\it 
PoornaPrajna Institute of Scientific Research,
4 Sadashivanagar, Bangalore 5600 80, India.\\
% E-Mail: sujit@physics.iisc.ernet.in\\
% Phone: 091 80 23612511/23619034, Fax: 091-80-2360-0228 \\
}
\date{\today}
    
%\pacs{74.78.Na}{Mesoscopic and nanoscale system}
%\pacs{74.81.Fa}{Josephson junction arrays and wire networks}
%\pacs{42.50.Lc}{Quantum Jumps.}

\date{\today}

\begin{abstract}
%\abstract{
We study the dissipation physics of one dimensional mesoscopic
superconducting quantum interference device array by using the 
field-theoretical renormalization group 
method. We observe length scale dependent superconductor-insulator 
quantum phase transition at very low temperature and also observe 
the dual behaviour
of the system for the higher and lower values of magnetic field. 
At a critical magnetic field, we also observe a critical behaviour 
where the resistance is 
independent of length.\\ 
\vskip .4 true cm
%\noindent Key Words: Field-theoretical renormalization group studies, 
%quantum phase transition, superconducting
%quantum interference devics
%\vskip .4 true cm
%\noindent PACS numbers: ~74.78.Na,
% 74.90+, 05.10.Cc 
Group methods
 
\end{abstract}
%\begin{document}
\maketitle

%\begin{multicols}{2}

%\section{ 1. Introduction}
1. Introduction: The transition at zero temperature or at very low 
temperature belongs to the class of phase transition
that is driven by the quantum fluctuatitons
of the system \cite{sub1}. The quantum fluctuations are controlled by the 
parameters of the systems such as the charging energies and Josephson 
couplings of the Josephson junctions array \cite{sond1}.
Here we present a field-theoretical renormalization group 
study to find the quantum dissipative phases of lumped
superconducting quantum interference device (Squid).
The quantum fluctuations of the system are controlled
by the externally applied magnetic field and $\alpha$
(= a ratio of quantum resistance to tunnel junction resistance).
In the Squid total current of device is modulated through the
applied magnetic flux. Therefore the total current in the Squid is    
$ I= 2 {I_c} sin( \theta ) 
| cos ( \frac{\pi {\Phi} }{{\Phi}_0} ) |$, where
$I_c$ is the critical current, $\Phi$ is the magnetic flux and
${{\Phi}_0} (= \frac{h}{2 e}) $ is the flux quantumi, $\theta$ is
the phase of superconducting order parameter.
Similarly the
Josephson coupling is also changing 
$ E_J~= 2 E_{J0} cos( \theta ) 
| cos ( \frac{\pi {\Phi} }{{\Phi}_0} )| $
, where $E_{J0}$ is the bare Josephson coupling.
So one can think that
a lumped Squid system (Fig. 1A) can be described in terms of an array 
of superconducting
quantum dots (SQD) but with a modulated Josephson 
coupling (Fig. 1B) and critical current \cite{havi1,havi2}.
This effective mapping will help us to analyze (analytically) the system
in detail. The experimentalist of Ref.3 and Ref.4 have also considered
the mesoscopic Squid system as a modulated Josephson junctions array.
We have been motivated from these well-accepted experimental findings
\cite{havi1,havi2}. 
The plan of this manuscript is the following.
Section (II) contains the
analytical derivations and the physical explanation for the occurrence 
of quantum dissipative phase in 
the lumped Squid
system. Conclusions
are presented in section (III).\\   
%______________________________________________________________
\section{2. Renormalization Group Study for Quantum
Dissipation Phase in the  Mesoscopic Lumped Squids}

We now present the basic dissipation physics of mesoscopic Squid
systems. 
The source of dissipation at very low temperature is due to the
appearance of phase slip centers.
Phase slip centers are of two kind: one is thermally 
activated phase slip center valid near to the superconducting
transition temperature, well described by LAMH theory \cite{tin1,amb1, hal1}, 
and the other,
quantum phase slip (QPS ) center occurs at T=0 K or at very low 
temperature due to
the quantum mechanical tunneling in different 
metastable states \cite{zai1}. 
The most important type of fluctuation which occurs during
this QPS process is that phase of the superconducting order parameter is
changing by $ \pm 2 \pi$ at a point in the system and the amplitude
of the order parameter vanishes at one point.
Appearence of QPS in different low dimensional superconducting 
systems is a common phenomena, we have cited only those references
which are related to our problem \cite{tin1,amb1,hal1,zai1}.
At first we prove the appearence of thermally activated phase slip
centres in lumped Squid system. We consider a Squid and use the 
Ginzburg-Landau theory to illustrate the physics of QPS. A
mesoscopic Squid is superconducting ring (nano scale size)
with two tunnel junctions.
We therefore use the cylindrical polar coordinates. 
The superconducting order parameter inside the
mesoscopic Squid is $ {\psi} (\phi) ~=~ {{\psi}_0} e^{i n {\phi} }$,
where $\phi$ is the azimuthal angle, $n$ is integer (winding number) 
and ${\psi}_0$ is constant. Here the Squid is pierced by 
the magnetic
flux $\Phi$. The vector potential along the tangential direction is
$ A_{\phi} ~=~ \frac{\Phi}{2 \pi R}$ ( $R$ is the radius of the
Squid). The free energy corresponding to this wave-function and
vector potential is
\beq
{F_s} (T)  =   {{F_s}}^{0} (T) ~+~ 
{V_s} \frac{{h^2}{| \psi |}^2 }{2 {m^*} R^2} {({\Phi} - n {{\Phi}_0})}^2 
 + \frac{c_1}{2 {\mu}_0} {{\Phi}^2} .
\eeq   
First term is the free energy contribution in absence of magnetic flux. 
$m^*$ is the effective mass of the quantum system and ${\mu}_0$ is the
permiability, $V_s$ is volume of the Squid, $c_1$ is a constant and the
last term represents the vacuum magnetic field energy. Free
energy is minimum when ${\Phi}~= n {\Phi}_0 $. Depending on the
externally applied magnetic flux, system is in one of the metastable
minima and a persistent current flows around the mesoscopic
Squid to maintain
the superconducting state. The system can jump from one metastable
minimum to the other minimum to lower the energy. This is the source of
dissipation and the decay of persistent current. Such an event
corresponds to a change in the winding number ($n$), hence it is a phase
slip. The rate of thermally activated phase slip is 
$ \frac{1}{{\tau}_1 } \sim e^{- \frac{e_b}{ {k_B} T} }$, where
${e_b}~\sim {V_s} \frac{{h^2}{| \psi |}^2 }{2 {m^*} R^2} $ 
is the barrier energy.
For the bulk Squid, $V_s$ is large and hence the appearance of phase slip
centers is quite unusual. But for the mesoscopic system of nano-scale
size such phase slip centers are likely possible. In the nano-wire,
there are many evidences for the presence of thermally activated
phase slip centers near the
superconducting transition temperature \cite{zai1}. 
The basic transport properties of low dimensional
tunnel junction systems and nano-wire have many similarities \cite{lar2,naza1}.
So we also expect the appearance of thermally activated phase slip
centers for mesoscopic lumped Squid system. Temperature
region close to the transition temperature 
of lumped Squid system \cite{havi1, havi2} is the 
region for thermally
activated phase slip centers. \\
Here we prove the appearance
of QPS in SQD array with modulated Josephson coupling 
through an analysis of a minimal model.
We consider two SQDs are separated by a Josephson junction.
These two SQDs are any arbitrary SQDs of the array. Appearance of
QPS is an intrinsic phenomenon (at any junction at any instant)
of the system. The
Hamiltonian of the system is 
\beq
H~=~ \sum_{i} ~\frac{{n_i}^2}{2 C}~-~ {E_J} | cos ( \frac{\pi \Phi}{{\Phi}_0} ) | \sum_{i} 
cos(~{\theta}_{i+1}~-~{\theta}_{i})
\eeq  
where $n_i$ and ${\theta}_i $ are respectively the Cooper pair density and
the superconducting phase of the i'th dot and $C$ is the capacitance
of the junction. 
The first term of the Hamiltonian present the Coulomb charging energies
between the dots and the second term is nothing but the Josephson
phase only term with modulated coupling, due to the presence of 
magnetic flux. We will see that this model is sufficient to capture 
the appearence of QPS in SQD systems. Hence this Hamiltonian has
sufficient merit to capture the low temperature dissipation physics
of SQD array.
In the continuum limit, 
the partition function of the system is given by, 
$
Z= \int  D \theta (x,\tau)  ~e^{- ~ {S_Q}  (x,\tau) },
$
where 
$
S_{Q} = \int d \tau 
\int dx \frac{E_J}{2} ~[{({\partial}_{\tau} \theta (x,\tau))}^2
 + {( {\partial}_x \theta(x,\tau) )}^2 ] 
$.
%$S_Q = \frac{E_J}{2} \int dx {({\partial}_{\mu} \theta (x))}^2 $
%(in covariant notation).
The action is quadratic in scalar-field $\theta (x,\tau )$, 
where $\theta (x,\tau )$
is a steady and differentiable field, so
one may think that no phase transition can occur
for this case. This situation changes drastically
in the presence of topological excitations for which $\theta (x,\tau)$ is
singular at the center of the topological excitations. So for this
type of system, we express the $\theta (x, \tau)$ into two components: 
$\theta (x, \tau) = {{\theta}_0} (x, \tau) + {{\theta}_1} (x, \tau)$, where 
${{\theta}_0} (x, \tau)$
is the contribution from attractive interaction of the system and 
${{\theta}_1} (x, \tau)$ is the singular part from topological excitations.     
We consider at any arbitrary time $\tau$, a topological excitation with
center at $X (\tau)~=~ ( {x_0} (\tau), {{\tau}_0} (\tau))$ 
. The angle measured
from the center of topological excitations between the spatial coordinate 
and the x-axis
$
{{\theta}_1} (x, \tau) ~=~ tan^{-1} (~\frac{{\tau}_0 - \tau}{x_0 -x} ) 
$
. The derivative of the angle is
$
{\nabla}_x {{\theta}_1} (x- X(\tau) )~=~ \frac{1}{{|x - X(\tau)|}^2}
 [-({\tau}_0 - \tau ), (x - x_0 )]
$
which has a singularity 
at the center of the topological excitation. 
Finally we get an interesting result when
we integrate along an arbitrary curve encircling
the topological excitations. 
$
\int_{C1} dx  {\nabla}_x  {{\theta}_1} (x - X(\tau) )~=~ 2 \pi . 
$
So we conclude from our analysis that when a topological
excitation is present in the SQD array, the phase difference
,$\theta$, across the junction of quantum dots jumps by 
an integer multiple of $2 \pi$. This topological excitation is
nothing but the QPS in the $(x,\tau)$ plane. 
According to the phase voltage Josephson relation
, $V_J~=~ \frac{-1}{2e} \frac{d \theta}{dt}$ ($t$ is the time),
a voltage drop occurs during this phase slip, which is the source
of dissipation. This analysis is valid when $\Phi$ is away from
$\frac{{\Phi}_0}{2}$. Otherwise system is in the superconducting
Coulomb blocked phase.   \\
Now the problem reduces to finding the quantum dissipation physics 
of SQD array with modulated Josephson couplings and critical current.
There are
a few interesting studies, following the prescription of
Calderia and Legget \cite{leg1},
to uncover the quantum dissipation physics of 
low dimensional tunnel junctions system \cite{sch1,su1,schon,fazi}.
Our starting quantum action is the same with Ref. \cite{sch1,su1,schon,fazi}.
We will see the scaling analysis of RG equations derive from this action is
sufficient to explain the experimental findings of Ref. [3-4].
\beq
S_1 ~=~ S_0 ~+~ \frac{\alpha}{4 \pi T}~ \sum_{m} {{\omega}_m}
{|{\theta}_m|}^2 .
\eeq
Here, $S_1$ is the standard action for the system with tilled wash-board
potential \cite{sch1,su1,schon,fazi} 
to describe the dissipative physics for low dimensional tunnel junction
systemi, $m$ is the Matsubara frequency.
Where 
$
S_0  = \int_{0}^{\beta} 
 \frac{C}{8 e^2} {(\frac{d \theta (\tau)} {d \tau})}^2 + V({\theta}{(\tau)})
$
,
$\alpha ~=~ \frac{R_Q}{R_s}$, Matsubara frequency 
${\omega}_m ~=~ \frac{2 \pi}{\beta} m$ and $R_Q$ 
($ = 6.45 k \Omega$) is the quantum resistance and $R_s$ is
the tunnel junction resistance, $\beta$ is the inverse temperature.
Here $ V ( \theta ) ~= -\frac{I_c}{2e} 
|cos(\frac{\pi \Phi}{{\Phi}_0} )| + \frac{ I \theta}{2 e}$. 
We would like to exploit the renormalization group (RG) calculation
for weak potential.  
Without loss of generality, we do the
analysis for $I~=~0$ case as finite $I$ only inclined the potential
profile. Since we are interested in the low energy excitations, we
can ignore the contribution of ${|{\omega}_m|}^2$ compare to
${|{\omega}_m|}$. So in $S_0$, we only consider the second term. 
\beq
S_0 ~=~ \frac{-I_c}{2e} |cos( \frac{\pi \Phi}{{\Phi}_0} )| 
\int_{0}^{\beta} cos (\theta) d {\tau}
= V_1 \int_{0}^{\beta} cos (\theta) d {\tau}
\eeq 
We can write the final action as
\beq
S_1 ~=~   \frac{\alpha}{4 \pi T} \sum_{m}
{|{\theta}_m|}^2 ~ + V_1 \int_{0}^{\beta} cos (\theta) d {\tau}
\eeq
The RG equation of the above action to study the low energy
excitations is the following:
\beq
\frac{d V_1}{d lnb}~=~ (1 - \frac{1}{\alpha}) V_1 .
\eeq
We have derived the above RG 
equation following the prescription
of Ref. \cite{shan1}. Here $b$ is a number ratio of the two energy
scale.
The time evolution
of the coupling constant is 
$ {V_1} (t) = {V_1} (0) e^{ (1 - 1/{\alpha} ) t}$. 
We are mainly interested in the low energy theory of the system,
suppose we consider the low energy
frequency as ${\omega}_m$ then the corresponding time is
$t~=~ ln (\frac{\Lambda}{{\omega}_m} )$. The coupling constant 
at maximum time reduces to 
$ V_1 = {V_1}(0) {(\frac{\Lambda}{{\omega}_m})}^{1- \frac{1}{\alpha}} $.
We consider the lowest frequency allowed by the 
Matsubara allowed frequency quantization, i.e., 
${\omega}_m ~= 2 \pi T$, so ${V_1} (T) ~\propto {T}^{ \frac{1}{\alpha} ~-~1 }$.
When we consider a finite system of length, $L$ ($ L > 1 $), then we might argue 
that the mode ${\theta} (k)$ is quantized with 
${\omega}_{1}~=~ \frac{\pi v}{L}$, 
where ${\omega}_1$ is the lowest frequency and $v$ is the velocity of
low energy excitations.
So the effective value of the
coupling at the lowest frequency is 
${V_1} (L) ~\propto {L}^{1 - \frac{1}{\alpha} }$.
We observe that the potential $V_1$ increases for $\alpha > 1$ and decreases
for $\alpha < 1$. So $ \alpha ~= 1 $ is the phase boundary. When 
$\alpha > 1$ owing to the strong dissipation effects particle comes to rest
at one of the minima of the potential (local or particle like character), 
i.e., the system is confined in one of the metastable current
carrying state. Hence the system is in the superconducting phase.
It is interesting to observe 
that dissipation favors to stabilize the superconducting phase,
$\alpha < 1$, implies weak dissipation 
has no effect on the potential. The
phase fluctuation is large around the dots (nonlocal or wave like process). 
As a result, there is
no phase coherent state in the system. 
Therefore there is no superconducting
phase.\\
We also observe from our study that the applied magnetic flux has
no effect on RG equation at the one-loop level. We consider
the effect of magnetic flux at the phenomenological level, i.e.,
we replace $\alpha$ by ${\alpha}' 
=~ {\alpha} |cos (\frac{ \pi \Phi}{{\Phi}_0} )| $.
When $\Phi$ is zero or an integer multiple of flux quantum,
the flux has no effect on the dissipation physics.
For the larger values of ${\Phi}$, (small ${\alpha}'$), make
the quantum fluctuations in the system 
large thereby destroying the phase coherence of states. So
the higher magnetic field drives the system from the superconducting
phase to the insulating phase. This is consistent with the experimental
findings \cite{havi1,havi2}.
The analytical structure of our derive RG equation (Eq. 6) has some
similarity with the RG equation of single impurity Luttinger liquid 
\cite{kane}.
But the initial Hamiltonians of these
two problems are quite different.\\   
In the strong potential, tunneling between the minima
of the potential is very small.  
In the imaginary time path integral formalism
tunneling effect can be described
in terms of instanton physics.
We will see that the strong coupling physics of our system can be described
interms of tunneling physics \cite{gia1,kane,furu}. 
%%%%%%%%%%%%%%%%%%%%%%%%%%%%%%%%%%%%%%%%%%%%%%%%%%%%%%%%%%%%%%%%%%%%%%%%%%%%
In the imaginary time path integral
formalism the potential is inverted  
and therefore the particle can not reside at the maximum of the 
potential for long time 
and rolls down to  one of the potential minima. It is convenient
to characterize the profile of $\theta$ in terms of its time derivative,  
\beq
\frac{d \theta {( \tau)} }{d {\tau} }~=~\sum_{i} e_i h (\tau - {\tau}_i),
\eeq
where  $h (\tau - {\tau}_i)$ is the time derivative at time $\tau$ of
one instanton configuration. 
${\tau}_i$ is the location of the ith instanton, $e_i = 1$ and $-1$
are respectively the topological charge of instanton and anti-instanton.
Integrated the function $h$ from 
$-\infty$ to $\infty$,
$ \int_{-\infty}^{\infty} d \tau h (\tau) = {\theta} (\infty)
- {\theta}({-\infty}) = 2 \pi . $ So the appearance of instanton
(anti-instanton) is  nothing but the appearance of QPS in the
lumped Squid system 
and it leads to the dissipative phase of the system.
Larkin $et~al.$ \cite{lar2} have also supported the idea of QPS as a appearance
of instanton (anti-instanton). 
So our system reduce to a neutral system consist of equal number of
instanton and anti-instanton.
Now our prime task is to present the partition function of the system.
After a few steps, we will come to that stage. One can find the expression
for ${\theta} ( \omega)$, after the fourier transform to the both sides
of Eq. 7 and that yield
\beq
{\theta} ( \omega ) = \frac{i}{\omega} \sum_{i} e_i h (i \omega) 
e^{i {\omega} {\tau}_1 }  
\eeq
Now we substitute this expression for ${\theta} ( \omega )$ in the
second term of Eq. 3, finally we get this term as
$ \sum_{ij} ~F({\tau}_i  - {\tau}_j ) {e_i} {e_j} $, where
$
F ({\tau}_i  - {\tau}_j ) = \frac{\pi \alpha}{ T}
\sum_{m }~ \frac{1}{{ |{\omega}_m}|}  
 \simeq  ln ({\tau}_i - {\tau}_j )
$ 
. We obtain this expression for very small values of 
${\omega}$ ( $\rightarrow 0 $).
So effectively $ F ({\tau}_i  - {\tau}_j )$ is representating the Coulomb
interaction between the instanton and anti-instanton. 
This term is the main source of dissipation physics
of the system. Following
the standard prescrption of imaginary time path integral formalism, we can
write the partition function of the system \cite{gia1,kane,furu}.    
\beq
Z ~=~ \sum_{n=0}^{\infty} {z}^{n} \sum_{e_i} \int_{0}^{\beta} d {\tau}_n
\int_{0}^{{\tau}_{n-1}}d {\tau}_{n-1} ...\int_{0}^{{\tau}_2} d {\tau}_1
e^{-F ({\tau}_i  - {\tau}_j ) {e_i}{e_j} }. 
\eeq    
Where $z^n$ is the contribution from instanton, $z~=~ e^{-S_{inst}} $
($ S_{inst} \simeq \sqrt{{V_1} C}$ ).
We may also write this expression as
\bea
Z & = & \sum_{n=0}^{\infty} {z}^{n} \sum_{e_i} \int_{0}^{\beta} d {\tau}_n
\int_{0}^{{\tau}_{n-1}}d {\tau}_{n-1} ...\int_{0}^{{\tau}_2} d {\tau}_1 \nonum\\
& & e^{ \sum_{m} \frac{{|{{\omega}_m}| }{|q ({{\omega}_m})|}^2}
{ (4 \pi { \alpha }^{'}  T) }
+ i \sum_{i} e_i {q_i} ({\tau}_i )}  . 
\eea    
Here $q_{\tau}$ is the auxillary field which arises during the functional
integral. We again introduce $ {\alpha}^{'} $ instead of
$\alpha$ to consider the effect of applied magnetic field.  
After extensive calculation, we get  
the partition function,
\beq
Z ~=~ \int D q (\tau) e^{-\frac{1}{4 \pi {\alpha}' T} \sum_{m} | \omega |
{|q_m|}^2  ~+~ 2 z \int_{0}^{\beta} d \tau cos( q( \tau ) ) }. 
\eeq    
$q ( \tau )$ is the auxiliary field. 
Following the method of previous paragraphs 
, we finally obtain the RG equation 
\beq
\frac{d z}{d lnb}~=~ (1 - {\alpha}') z
\eeq
The analytical
structure of the quantum action and the RG equation is the same with
weak potential, it implies the following mappings 
$ {\alpha}' \leftrightarrow \frac{1}{{\alpha}'} $ and 
$ V_1  \leftrightarrow z$. Hence there is a duality 
in this problem between the weak and strong potential. 
One can also find the analytical expression for the variation of 
$z $ as a function of temperature and length as 
we derive for weak coupling case, the only change being ${\alpha}'$
replaced by $1/{\alpha}'$. 
Fugacity depends 
on temperature and length scale as, 
$ z(T)~\propto T^{{\alpha}' -1} , z(L)~\propto L^{1 - {\alpha}'}$.\\
In this complicated system, we estimate
the behavior of resistance from the behavior
of fugacity. It is expected to 
scale as $ z^2 $, 
(because the major contribution of voltage/resistance
occurs from the second order expansion of partition function, i.e.,
from the square of fugacity).  
In our study, resistance is evolving due to dissipation
effect at very, low 
temperature (less than the superconducting Coulomb blocked temperature).
According to our calculations, for large dissipation ($ {\alpha}' > 1$), 
$ R(T) \propto  R_Q {T^{{\beta}_1}}$, ${{\beta}_1} > 0$.
Therefore at very low temperature, the system shows
the superconducting behavior. When ${\alpha}' < 1$,
the resistance of system
$ R(T) \propto  R_Q  {T^{{- \beta}_2}}$, 
${{\beta}_2} > 0 $. So at very low temperature,
the resistance of the system shows Kondo-like divergence behavior
and the system is in the insulating phase.
According to our calculations, for large dissipation (${\alpha}' > 1$), 
$ R(T) \propto R_Q {L^{- {\gamma}_1}}$, ${{\gamma}_1} > 0$.
Therefore the longer array system shows
the less resistive state than shorter array in the 
superconducting phase of the system. When ${\alpha}' < 1$,
the resistance of system
$ R(T) \propto R_Q {L^{{ \gamma}_2}}$, 
where ${{\gamma}_2} > 0 $ 
($ {{\beta}_1}, {{\beta}_2}, {{\gamma}_1}$ and ${{\gamma}_2}$
are independent numbers).
So the resistance at the insulating state is larger for longer 
array system than the shorter one.
So we find the dual behavior of the resistance (voltage)
for lower and higher values of magnetic field.  
When ${\alpha}' = 1$, i.e.,
$ {\Phi} ~=({{\Phi}_0}/{\pi})cos^{-1} (\frac{1}{\alpha}) $,
the system has no length scale dependence superconductor-insulator
transition at very low temperature. This is the critical
behavior of system for a specific value of magnetic field.
These theoretical findings are consistent with the experimental
observations \cite{havi1,havi2}.
Fig. 2 shows the variation of resistance with temperature. At
higher temperature, larger than $T_c$. SQD array system is in the
normal phase and the tunneling between the dots is
the sequential tunneling (one after another, tunneling of Cooper pairs). 
We have shown in Ref. \cite{sar2}, that
superconductivity occurs due to the co-tunneling effect
(higher order tunneling of Cooper pairs, virtual process). A presence
of finite resistance at the superconducting phase
(between $T_1$ and $T_2$ )
due to the dissipation effect and also for the
presence of finite tunneling conductance
(i.e, the finite resistance).
Low resistance superconducting phase or insulating phase occurs at
very low temperature, smaller than the superconducting Coulomb blocked
temperature.
\\
Low resistance superconducting phase or insulating phase occurs at
very low temperature (it will occur for few mili-kelvin), 
smaller than the superconducting Coulomb blocked
temperature. In experiment, they have measured up to 50 mK, so they have not
found the decaying tendency of resistance at very 
low temperature.
We have not considered the the classical phase ($ E_J >> E_C$) 
of the system. In this phase, one can also obtain the dissipative
phase with phase slip centers, but one can loses the informations
of intermediate quantum phases for the system \cite{sar2}. 
Chakarvarty $et~ al.$ \cite{su1} and
Larkin $et~al.$ \cite{lar2} had studied QPS for the classical phase.
Currently Fistual $et~al.$ \cite{fis} have done some interesting work on collective
Cooper pair transport in the insulating state of one and two dimensional
Josephson junctions array. They have studied the current-voltage characteristics
revealing thermally activated conductivity at small voltages and threshold
voltage depinning. Our analytical approach to study the one dimensional
mesoscopic squid system is quite different from them \cite{fis}.  
\\
We have studied the clean system. Here we explain the effect of
impurities in superconducting dots or in the tunnel barrier. Nonmagnetic
impurities were found not to affect the Josephson supercurrent. This is in 
agreement with Anderson's theorem \cite{and} of dirty superconductor. In presence
of paramagnetic impurities, there is an exchange interaction between
the spin of conduction electrons with the magnetic impurity spin. The
spin of the magnetic impurity polarize the spin of electrons and 
interferes with their tendency for pair formation in the singlet
state. As one increase the probablity ($\Gamma$) of scattering with
spin flip. For highervalues of $\Gamma$, the system enters into the
gapless region and above a critical value
( ${{\Gamma}_c} = \frac{{\Delta (0,0)}}{2}$, ${\Delta (0,0)}$ being 
the order parameter for the superconductor at zero temperature in
absence of impurity. ) the superconductivity is destroyed \cite{bar}. For such
situation, our system is in the Coulomb blocked insulating
phase. The Josephson supercurrent
is also zero for a clean dot when the applied magnetic flux of the
system is half-integer multiple of flux quantum. At those values of
magnetic flux, our system is in the Coulomb blocked insulating phase.  
\section{ 3. Conclusions}
We have observed the length scale dependent superconductor-insulator
quantum phase transition and the dual behaviour of the system for
smaller and larger values of magnetic field, in a one dimensional
mesoscopic lumped Squid systems. We have also observed
a critical behaviour where the resistance is independent of length
at a critical magnetic field.   
We find that weak and strong potential results
are self dual. Our theoretical findings have  
experimental relevance of lumped Squids system \cite{havi1,havi2}.    

The author would like to acknowledge Prof. T. Giamarchi for useful 
discussion during the final formating of the manuscript and also 
Physics Department of IISc for extended facility.
The author would like to thanks Dr. N. Mahajan and Dr. B. Mukhopadhyay 
for reading the
manuscript very critically. Finally the author thanks Prof. N. Kumar
for his interest in this work.

\begin{figure} 
\includegraphics[scale=0.35,angle=0]{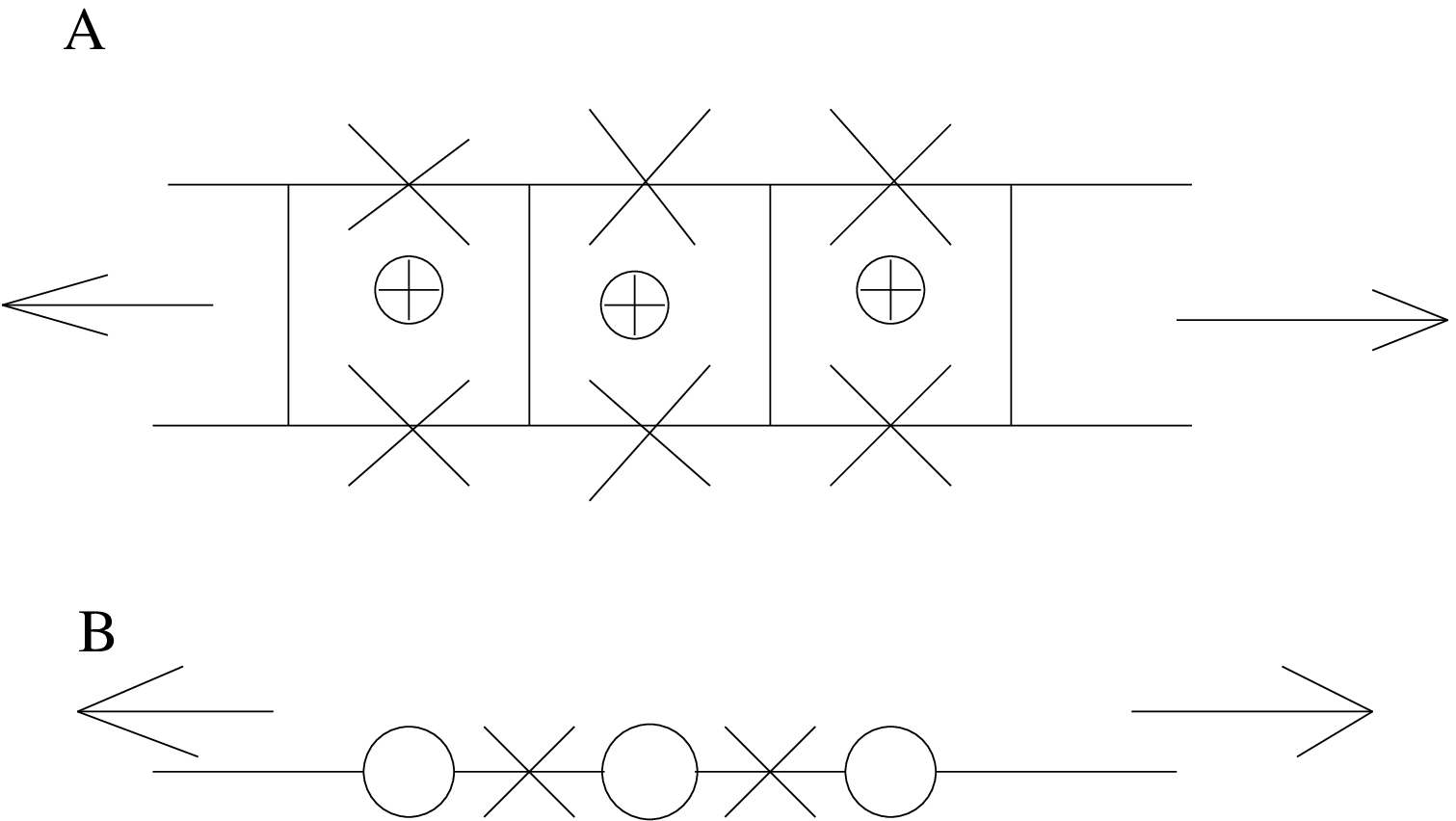} 
\caption{A. Schematic diagram of one dimensional array of
small capacitance dc Squids. Each plaque is a squid with
two Josephson junctions marked by the cross.
Circle with plus sign represent the applied magnetic flux $\Phi$. 
B. Equivalent representation of system A, where the dots are
connected through tunnel junctions and the Josephson couplings
of this system is tunable due to presence of magnetic flux $\Phi$.}
\label{Fig. 1 }
\end{figure}
%\end{thebibliography}
\begin{figure}
\includegraphics[scale=0.50,angle=0]{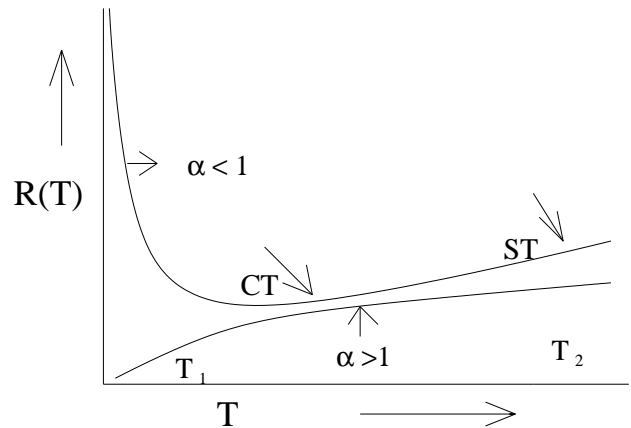}
\caption{ Schematic phase diagram, variation of resistance
with temperature. ST (see the text) is sequential tunneling regime 
and CT (see the text) is
the Co-tunneling regime. $T_2$ (${\sim}$ few Kelvin) is
higher than superconducting
transition temperature
and $T_1$ (${\sim}$ few milli Kelvin )
is the temperature region much less than the superconducting
Coulomb blocked transition temperature.}
\label{Fig. 2 }
\end{figure}


\begin{references}
%\begin{thebibliography}{0}
\bibitem{sub1} Subir Sachdev, {\it Quantum Phase Transition} (Cambridge
University Press Cambridge, 1998)

\bibitem{sond1} S. L. Sondhi., $et al.$, Rev. Mod. Phys. {\bf 69},
(1997) 315.

\bibitem{havi1} E. Chow, P. Delsing, and D. B. Haviland, Phys. Rev. Lett,
{\bf 81}, (1998) 204.

\bibitem{havi2} M. Watanbe and  D. B. Haviland, Josephson Structures and
Superconducting Electronics, {\bf 6} edited by A. V. Narlikar
(Nova Science Publishers, New York, 2002).

\bibitem{tin1} M. Tinkham in {\it Introduction to  Superconductivity} 
(McGRAW-HILL, New York 1996).

\bibitem{amb1} J. S. Langer and V. Ambegaokar, Phys. Rev, {\bf 164}
(1967) 498.

\bibitem{hal1} D. E. MaCumber and  B. I. Halperin, Phys. Rev. B, {\bf 1}
(1970) 1054.

\bibitem{zai1}  D. S. Golubev and A. D. Zaikin, Phys. Rev. B, {\bf 64}
(2001) 014504; A. D. Zaikin $et~al.$, Phys. Rev. Lett., {\bf 78}
(1997) 1552.

\bibitem{lar2} K. A. Matveev,  A. I. Larkin and  L. I. Glazmann, 
Phys. Rev. Lett. {\bf 89}, (2002) 096892.

\bibitem{naza1} J. E. Mooij and Yu. V. Nazarov, Naturephys, {\bf 2}, (2006) 169.

\bibitem{leg1} A. O. Calderia and A. J. Leggett, Phys. Rev. Lett,
{\bf 46}, (1981) 211; Ann. Phys {\bf 149}, (1983) 374.

\bibitem{sch1} A. Schmid, Phys. Rev. Lett. {\bf 51}, (1983) 1506.

\bibitem{su1} S. Chakravarty $et~al.$, Phys. Rev. B {\bf 37}, (1988) 3238
; S. Chakravarty, Phys. Rev. Lett. {\bf 49}, (1082) 681.

\bibitem{schon} S. Schon  and  A. D. Zaikin, Phys. Reports {\bf 198},
(1990) 237.

\bibitem{fazi} R. Fazio, and H. van der Zant, Physics Report
{\bf 355}, (2001) 235.

\bibitem{shan1} R. Shankar, Rev. Mod. Phys {\bf 66}, (1994) 129.

\bibitem{gia1} Giamarchi. T in {\it Quantum Physics in One Dimension} 
(Clarendon Press, Oxford, 2004).

\bibitem{kane} C. Kane and M. P. A Fisher, Phys. Rev. B {\bf 46},
(1992) 15233; C. Kane and  M. P. A Fisher, Phys. Rev. Lett {\bf 68},
(1992) 1220. 

\bibitem{furu} A. Furusaki and N. Nagaosa, Phys. Rev. B {\bf 47}, (1993) 4631
; ibid Phys. Rev. B {\bf 47}, (1993) 3827.

\bibitem{sar2} S. Sarkar, Europhys. Lett, {\bf 76}, (2006) 1172.

\bibitem{fis} M. V. Fistul,  V. M. Vinokur and T. I. Batrunia,
cond-mat/0708.2334 .

\bibitem{and} P. W. Anderson, J. Phys. Chem. Solids {\bf 11}, (1959) 26.

\bibitem{bar} A. Barone and G. Paterno in {\it Physics and Application of
the Josephson Effect} (USA, John Wiley and Sons, 1982).  


\end{references}
\end{document}